\documentclass[doublecol,linenumbers]{epl2} 
\RequirePackage{amssymb,amsmath}
\usepackage{indentfirst}
\usepackage{color}
\newcommand{\be}{\begin{equation}}
\newcommand{\ee}{\end{equation}}

\newcommand{\mr}{\mathring}

\newcommand{\ba}{\begin{eqnarray}}
\newcommand{\ea}{\end{eqnarray}}

\usepackage{color}
\title{Thick Braneworlds and the Gibbons-Kallosh-Linde No-go Theorem in the Gauss-Bonnet Framework}
\shorttitle{Thick Braneworlds  in the Gauss-Bonnet Framework} 

\author{M. Dias\inst{1}  J. M. Hoff da Silva\inst{2} \and Rold\~ao da Rocha\inst{3,4}}

\institute{      
\inst{1} Departamento de Ci\^encias Exatas e da Terra, Universidade Federal de S\~ao Paulo
Diadema, SP, Brazil.\\              
  \inst{2} Departamento de F\1sica e Qu\1mica, Universidade
Estadual Paulista, Av. Dr. Ariberto Pereira da Cunha, 333, Guaratinguet\'a, SP,
Brazil.\\
  \inst{3} Centro de Matem\'atica, Computa\c c\~ao e Cogni\c c\~ao, Universidade Federal do ABC 09210-170, Santo Andr\'e, SP, Brazil.\\
  \inst{4} International School for Advanced Studies (SISSA),
Via Bonomea 265, 34136 Trieste, Italy.
}

\abstract{
The sum rules related to thick braneworlds are constructed, in   order to encompass Gauss-Bonnet terms. The generation of thick branes is hence proposed  {} {in a periodic extra dimension scenario}, what circumvents the Gibbons-Kallosh-Linde no-go theorem in this context. }

\begin{document}

\maketitle

\section{Introduction}

Braneworld models {have been providing prominent}  aspects concerning  high energy physics, since the seminal works on warped geometry \cite{HW,RS,RS1,Rey,Cvetic01,anto} and some ramifications   \cite{GOG}. 
The warped structure {provides a solution} for the hierarchy problem. For instance, the standard model in \cite{RS,RS1}  presents two branes placed at the singular points of a $S^1/\mathbb{Z}_2$ orbifold. Some significant generalizations  were further proposed, leading to  thick brane solutions \cite{GREMM}.

Physical constraints arising {in the context of} braneworld models \cite{ELL}  provide a framework that correlates physical {} {information} to mathematical conditions, in order to {} {specify} a viable braneworld {} {structure} \cite{KGL}. To employ the  braneworlds sum rules \cite{KGL}, the existence of a compact internal space without boundary  is demanded, which in 5D is provided by the above mentioned orbifold. With respect to this geometry, thick generalizations of the two-brane Randall-Sundrum and Kogan-Mouslopoulos-Papazoglou-Ross-Santiago solutions \cite{KM}, without singular sources, were shown to be precluded. {} {It is precisely the} Gibbons-Kallosh-Linde no-go theorem, {} {asserting that} the integral  of a quadratic quantity along the extra dimension equals zero. 
\par 
Gravity in four dimensions occupies an important place when compared to higher dimensional ones. Indeed, although the Einstein-Hilbert action is unique in 4D, it does hold on higher dimensions. In 5D, Lovelock's theorem  asserts that the most general unique action can be achieved by including the Gauss-Bonnet term  to the usual Einstein-Hilbert action, added by the cosmological constant \cite{Lovelock}. The Gauss-Bonnet Lagrangian appears moreover in an effective action approach to string theory, being equivalent to the leading order quantum correction to gravity \cite{gross}.  Besides, the Gauss-Bonnet term is the only curvature term that provides  ghost-free self-interactions for the graviton, corresponding to the no ghost  character of string theories \cite{zwiebach}. The Gauss-Bonnet terms moreover   stabilizes the extra dimension in KK models \cite{madore2}, further providing exact solutions in Gauss-Bonnet theories \cite{w1, deser}. 
General bulk solutions and their induced braneworld  cosmology were thus comprehensively studied in \cite{Charmousis:2002rc, Binetruy:2002ck}. The late-time cosmic acceleration and the transition from 
the deceleration to acceleration can be mimicked as well \cite{nojiri1, cogn}. 
Black hole thermodynamics has been also studied in the context of Gauss-Bonnet gravity \cite{Cai:2001dz, deser}. 
\par
Since the Gauss-Bonnet term can break energy conditions, it plays a role of a singular source, as for instance the negative tension brane in Randall-Sundrum models \cite{RS,RS1}.
In the context of higher-dimensional models, similar theorems have been proved. For example, the Maldacena-Nunez no-go theorem, that forbids compactification in an accelerating universe \cite{Maldacena:2000mw}.

The aim of this paper is to show that  it is possible to circumvent the Gibbons-Kallosh-Linde theorem hypothesis, by taking into account  gravity in the bulk.
Indeed, terms coming from the extended setup of gravity can alleviate the strong constraint imposed by the sum rules. The necessity of a generalized gravitational theory is based, in fact, on the theorem itself. In fact, the theorem does not take into account an specific type of scalar field potential. 
 
 We shall thus consider the next order term of the Lovelock series \cite{Lovelock}, namely, the Gauss-Bonnet term. By accomplishing it,  the  framework to be  analyzed shall not perform a complete model. In fact, some important ingredients in the sum rule formalism  -- the so called partial traces -- do not provide closed relations. It is worth though, since by assuming that the contribution coming from the Gauss-Bonnet term is weak in the bulk, we are able to show the possibility of a generalization of the Randall-Sundrum  setup to thick braneworlds,  avoiding the necessity of a negative brane tension. 

This paper is organized as follows: in Sec. II we accommodate the  formalism concerning the sum rules, in order to encompass the additional terms coming from the Gauss-Bonnet geometry in a given context. The application to the Gibbons-Kallosh-Linde theorem is performed in Sec. III. In Sec. IV we point out the conclusion and final remarks. 

\section{The Framework}

The braneworld sum rules were developed in such a way that the gravitational dynamical equation, even without being solved, can be used to extract necessary conditions that enable a physical setup. The framework is constructed upon geometrical considerations, being useful in dealing with compact internal spaces, as the one regarding  an orbifold extra dimension, within an warped spacetime. 
The general formalism is provided by considering that the bulk is $D$-dimensional, whilst the $p$-brane has $(p+1)$ dimensions. The internal space has thus  $(D-p-1)$ compact dimensions. The metric $G_{MN}dx^Mdx^N$ is given by the general expression 
\be 
ds^2\!=\!W^2(y)g_{\mu\nu}dx^\mu dx^\nu+g_{mn}(y)dy^m dy^n\,\label{metrica}
\ee where $M,N=0,\ldots, D-1$ are bulk indexes, $\mu,\nu=0,\ldots,p$ are braneworld indexes and $m,n= p+1,\ldots,D-1$ stand for indexes related to the internal space. Moreover, $W^2(y)$ denotes the warp factor,  $g_{\mu\nu}$ and $g_{mn}(y)$ are respectively the brane and the internal space metric components.

The warp factor, the brane, and  the scalars of curvature related to the  internal space as well ($\,^{(p+1)}\!R$ and $^{(D-p-1)}\!R$, respectively) are related to the partial traces by the expression 
\begin{eqnarray}
&&\!\!\!\!\!\!\!\!\!\nabla\cdot(W^\alpha\nabla W)=\frac{W^{\alpha+1}}{p(p+1)}\Bigg\{\alpha\Big[\,^{(p+1)}\!RW^{-2}+R^\mu_{\;\mu}\Big]\nonumber\\&&+(p-\alpha)\Big[\,^{(D-p-1)}\!R-R^m_{\;m}\Big]\Bigg\},\label{2}
\end{eqnarray} 
where $R^\mu_{\;\mu}=W^{-2}g^{\mu\nu}\,\mr{R}_{\mu\nu}$ and $R^m_{\;m}=g^{mn}\,\mr{R}_{mn}$ are the partial traces. The sum results in the bulk scalar of curvature $\mr{R}=R^\mu_{\;\mu}+R^m_{\;m}$. A similar approach to a brane with torsion was considered in \cite{torsion}.  In Eq. (\ref{2}), $\alpha$ is a parameter that provides a condition for  the respective model. By  integrating with respect to the internal space, it is possible to find an 1-parameter ($\alpha$) family of consistency conditions. 
Let us hence implement the generalization concerning the Gauss-Bonnet term. As the partial traces are obtained from the brane and internal space Ricci tensors as well, the connection between the geometrical approach and the physical setup is performed by the Einstein equations.

Hereupon we denote by $\mr{A}$ all quantities that refer to $D$-dimensional objects. By considering the Gauss-Bonnet term
\begin{eqnarray}\label{GB} \!\!\!\!GB\!\!&\!:=\!&\!\!\!\!\mr{R}_{M\!N\!A\!B}\mr{R}^{M\!N\!A\!B}\!-\!4\mr{R}_{AB}\mr{R}^{AB}\!+\mr{R}^2\end{eqnarray} the Einstein equations reads \begin{eqnarray}
\mr{R}^M_{\;\;N}\!-\!\left.\frac{1}{2}G^M_{\;\;N}\,\mr{R}\!+\!2\alpha_2\Bigg(\!\!\,\mr{R}^M_{\;\,S\!K\!P}\,\mr{R}_N^{\;S\!K\!P}\!\!-\!2\,\mr{R}^M_{\;SNP}\mr{R}^{SP}\right.\nonumber\\\hspace*{-0.3cm}\!\!\!\!\!\!\!-2\,\mr{R}^M_{\;\;S}\,\mr{R}^S_{\;\;N}\!+\!\mr{R}\,\mr{R}^{M}_{\;N}\!\Bigg)\!-\! \frac{\alpha_2}{2}G^M_{\;N}GB=8\pi G_D T^M_{\;N},\label{3}
\end{eqnarray} 
where $G_D$ stands for the bulk gravitational constant and $\alpha_2$ provides the strength of the Gauss-Bonnet term contribution. In the computation of the partial traces,  the result should be finally expressed  in terms of the  stress tensor. Hence the contraction of Eq. (\ref{3}) imply that $(2-D) \mr{R}={16\pi G_D}\,T+{(D-4)}\alpha_2 GB$.  Now, by isolating $\alpha_2GB$ from the above equation  and reinserting it back into Eq. (\ref{3}) we obtain
\begin{eqnarray}
\hspace*{-0.3cm}\!\!\!\!\!\!\mr{R}_{M\!N}\!-\!G_{M\!N}\frac{\mr{R}}{4\!-\!D}\!+\!2\alpha_2\Bigg(\!\mr{R}_{M\!A\!B\!C}\mr{R}_N^{\;\,A\!B\!C}\!\!-2\mr{R}_{M\!A}\mr{R}^A_{\;N}\nonumber\\-2\,\mr{R}_{M\!A\!N\!C}\mr{R}^{AC}\!\!+\!\mr{R}\mr{R}_{M\!N}\!\Bigg)\!\!=\!8\pi G_D\!\Bigg(\!T_{M\!N}\!+\!\frac{G_{M\!N}}{4\!-\!D}T\Bigg).\hspace*{-0.3cm}\label{5}
\end{eqnarray}
Notice that in the limit $\alpha_2\rightarrow 0$ we have $\,\mr{R}=\frac{16\pi G_D}{2-D}T$, and Eq.(\ref{5}) is led to the $D$-dimensional Einstein equation, as expected. 

From Eq.(\ref{5}) the partial traces can be evinced. Starting from $R^\mu_{\;\mu}$ and writing the partial traces into Eq. (\ref{2}), it yields 
\begin{widetext}
\begin{eqnarray}
R^\mu_{\;\mu}&=&\left.\frac{1}{1+2\alpha_2\,\mr{R}}\Bigg\{\frac{8\pi G_D}{D-4}\Bigg((D-p-5)T^{\mu}_{\;\mu}-(p+1)T^m_{\;m}\Bigg)-\frac{(p+1)}{D-4}\,\mr{R}\right. \nonumber\\&-&\left.2\alpha_2 W^{-2}\Bigg[\,^{(p+1)}\!R_{\mu\alpha\beta\gamma}\,^{(p+1)}\!R^{\mu\alpha\beta\gamma}-
4\,^{(p+1)}\!R_{\mu\alpha}\,^{(p+1)}\!R^{\mu\alpha}+3\,\mr{R}_{\mu a\beta\gamma}\,\mr{R}^{\mu a\beta\gamma}-2\,\mr{R}_{\mu a}\,\mr{R}^{\mu a}\right. \nonumber\\&+&\left.3\,\mr{R}_{\mu ab\gamma}\,\mr{R}^{\mu ab\gamma}+\,\mr{R}_{\mu abm}\,\mr{R}^{\mu abm}-4\,\mr{R}^{\mu}_{\;a\mu\nu}\,\mr{R}^{a\nu}-2\,\mr{R}^{\mu}_{\;a\mu m}\,\mr{R}^{am}\Bigg] \Bigg\},\right. \nonumber\\R^m_{\;m}&=&\left. \frac{1}{1+2\alpha_2\,\mr{R}}\Bigg\{\frac{8\pi G_D}{D-4}\Bigg((p-3)T^{m}_{\;m}-(D-p-1)T^\mu_{\;\mu}\Bigg)-\frac{(D-p-1)}{D-4}\,\mr{R}\right. \nonumber\\&-&\left.\!\!2\alpha_2\Bigg[\,^{(D-p-1)}\!R_{mabn}\,^{(D-p-1)}\!R^{mabn}\!-\!4\,^{(D-p-1)}\!R_{mn}\,^{(D-p-1)}\!R^{mn}\!+\!3\,\mr{R}_{mab\gamma}\,\mr{R}^{mab\gamma}\!-\!2\,\mr{R}_{m\mu}\,\mr{R}^{m\mu}\right. \nonumber\\&+&\left.3\,^{(D)}R_{ma\beta\gamma}\,^{(D)}R^{ma\beta\gamma}+
\,^{(D)}R_{m\alpha\beta\gamma}\,^{(D)}R^{m\alpha\beta\gamma}-4\,\mr{R}^m_{\;am\gamma}\,\mr{R}^{a\gamma} -2\,\mr{R}^m_{\;\alpha m\gamma}\,\mr{R}^{\alpha\gamma}\Bigg]\Bigg\}. \right. \label{9}
\end{eqnarray}
\end{widetext}\begin{floatequation}
\mbox{\textit{see Eq. ~\eqref{9}}}
\end{floatequation}

 Some relevant points concerning the partial traces are worth to be mentioned. In fact, since $T=T^{\mu}_{\;\mu}+T^m_{\;m}$, then   the relation $R^{\mu}_{\;\mu}+R^m_{\;m}=\,\mr{R}$ holds.  Eqs. (\ref{9})  can moreover provide a trivial meaning to the word ``closed''. Indeed, since $\,\mr{R}$ is the sum of the partial traces, both expressions for $R^{\mu}_{\;\mu}$ and $R^m_{\;m}$ are written  iteratively in terms of both $R^{\mu}_{\;\mu}$ and $R^m_{\;m}$, among several other quantities. On the other hand, the generalization of Eq. (\ref{2}) in order to encompass the Gauss-Bonnet geometry is an extremely hard task. Nevertheless, it is still possible to provide physical information about the braneworld model. For instance, the  bulk scalar of curvature can be introduced by hand in the model  \cite{ELL,KGL,COMP}. 

 It is worth to emphasize that in the scope of string theory, the Gauss-Bonnet term is the next leading order term of the string tension expansion \cite{TORII,GROSS}. Thus, it is possible to expand the $\alpha_2$ factors disregarding the nonlinear contributions, namely, $(1+2\alpha_2\,\mr{R})^{-1}\simeq 1-2\alpha_2\,\mr{R}.$ Hence Eq. (\ref{2}), together with the $\alpha_2$-linear version of Eqs. (\ref{9}), yields
 \begin{eqnarray}
\nabla\cdot(W^\alpha\nabla W)&=&\left.\frac{W^{\alpha+1}}{p(p+1)}\Bigg\{\alpha\,^{(p+1)}\!RW^{-2}\right.\nonumber\\\quad+(p\!-\!\alpha)^{(D\!-\!p\!-\!1)}\!R\!\!\!\!\!\!&& \left.\!\!\!\!\!\!\!\!\!\!\!\! \,+\gamma_3\,\mr{R}+\!8\pi G_D\bigg(\tilde{\gamma}_1T^\mu_{\;\mu}+\tilde{\gamma}_2T^m_{\;m}\bigg)\right. \nonumber\\+2\alpha_2\bigg[\alpha W^{-2}&&\left.\!\!\!\!\!\!\!\!\!\!\!\!\!\!\!\!\!A_{GB}\!-\!\gamma_3\,\mr{R}+\!(p\!-\!\alpha)B_{GB}\bigg]\!\Bigg\} \right. ,\label{10}
\end{eqnarray} where $\tilde{\gamma}_1=\frac{p(D-p-1)-2\alpha(D-p-3)}{(D-4)}$, $\tilde{\gamma}_2=\frac{2\alpha(p-1)-p(p-3)}{(D-4)}$, and $\gamma_3=\frac{\alpha(p+1)+(p-\alpha)(D-p-1)}{(D-4)}$ are  coefficients depending on the dimensions of the bulk and upon the $p$-brane as well. The terms $A_{GB}$ and $B_{GB}$ are purely Gauss-Bonnet corrections provided respectively by 
\begin{eqnarray}
A_{GB}:= \!\,^{(p+1)}\!R_{\mu\alpha\beta\gamma}\,&&\left.\!\!\!\!\!\!\!\!\!\!\!\!\!\!\! ^{(p+1)}\!R^{\mu\alpha\beta\gamma}\!-\!
4\,^{(p+1)}\!R_{\mu\nu}\,^{(p+1)}\!R^{\mu\nu}\right.\nonumber\\-4\,\mr{R}^\mu_{\;a\mu\nu}\,\mr{R}^{a\nu}&&\left.\hspace{-0.7cm}-2\,\mr{R}^\mu_{\;a\mu m}\,\mr{R}^{am}\!\!\,+\mr{R}_{abc\mu}\,\mr{R}^{abc\mu}\!\!\right.\nonumber\\\!\!\!\!\!\!\!\!\!\!\!\!+3\,\mr{R}_{ab\mu\nu}\,\mr{R}^{ab\mu\nu}&&\!\!\!\!\!\!\!\!\!\!+
3\,\mr{R}_{a\alpha\beta\gamma}\,\mr{R}^{a\alpha\beta\gamma}\!-\!2\,\mr{R}_{a\mu}\,\mr{R}^{a\mu}\label{11}\end{eqnarray}and
\begin{eqnarray}
B_{GB}\!\!\!&:=&\left.\!\!\!\!\!\! \,^{(D-p-1)}\!R_{abmn}\,^{(D-p-1)}\!R^{abmn}\right.\nonumber\\&&\left.-
4\,^{(D-p-1)}\!R_{ab}\,^{(D-p-1)}\!R^{ab}-4\,\mr{R}^m_{\;am\gamma}\,\mr{R}^{a\gamma}\right.\nonumber\\&&\left.-2\,\mr{R}_{a\mu}\,\mr{R}^{a\mu}-2\,\mr{R}^m_{\;\alpha m\gamma}\!\!\,\mr{R}^{\alpha\gamma}+\,\mr{R}_{a\alpha\beta\gamma}\,\mr{R}^{a\alpha\beta\gamma}\right. \nonumber\\&&\left.+\,3\,\mr{R}_{ab\mu\nu}\,\mr{R}^{ab\mu\nu}+
3\,\mr{R}_{abc\mu}\,\mr{R}^{abc\mu}.\right.\!\!\label{12}
\end{eqnarray} 

Apart from the $A_{GB}$ and $B_{GB}$ contributions, Eq. (\ref{10}) is given in terms of source factors (stress-tensor terms) and moreover in terms of the brane, the internal space, and the bulk scalar of curvature as well. Within this expression we can study several physical possibilities by looking at these inputs, or setting its behavior for a given model.

After constructing the functional form of Eq. (\ref{10}),  the stress tensor can be taken into account, explaining the physical content of the model. Consider thus the following expressions: 
\begin{eqnarray}
-\frac{T^\mu_{\;\mu}}{(p+1)}\!\!\!\!&\!=\!&\!\Bigg[a-\sum_iT_q^{(i)}\tilde\Delta(y-y_i)\Bigg]+\tau^\mu_{\;\mu},\label{13}\\
\hspace*{-.9cm}\frac{T^m_{\;m}}{(1\!-\!D\!+\!p)}\!&\!=\!&\!\!\!-a\!-\!\sum_i(q\!-\!p)T_q^{(i)}\tilde\Delta(y\!-\!y_i)\!+\!\tau^m_{\;m}\,,\label{14}
\end{eqnarray} where $a:=\frac{\Lambda}{8\pi G_D}$ and $\tilde\Delta= \Delta^{(D-q-1)}$. In the above equations, $\Lambda$ stands for the bulk cosmological constant, $T_q^{(i)}$ is the tension of the $i^{\rm th}$ $q$-brane placed at $y_i$. Besides, $\Delta^{(D-q-1)}$ is the generalized delta term, to fix the brane position  \cite{COMP}, and the $\tau$ terms encompass additional fields in the bulk, related to the scalar field generating the thick brane. By substituting Eqs. (\ref{13}) and (\ref{14}) into (\ref{10}) it follows that
\begin{widetext}
\vspace{-0.6cm}\begin{eqnarray}
\hspace*{-1cm}\nabla\cdot (W^\alpha \nabla W)=\left.\frac{W^{(\alpha+1)}}{p(p+1)}\Bigg\{\alpha\,^{(p+1)}\!RW^{-2}+(p-\alpha)\,^{(D-p-1)}\!R+\gamma_3\,\mr{R}+8\pi G_D(1-2\alpha_2)\times\Bigg[\!-\!\frac{\Lambda}{8\pi G_D}\gamma_1\right.\nonumber\\\left.\!-\!\gamma_2\sum_{i}T_q^{(i)}\Delta^{(D-p-1)}(y-y_i)+\tilde{\gamma}_1\tau_{\mu}^{\;\mu}+\tilde{\gamma}_2\tau^m_{\;m}\Bigg]\!+\!2\alpha_2\Bigg[\!-\!\gamma_3\mr{R}+\alpha W^{-2}A_{GB}\!+\!(p\!-\!\alpha)B_{GB}\Bigg]\Bigg\},\right.\label{14}
\end{eqnarray}
\end{widetext}\begin{floatequation}
\mbox{\textit{see Eq. ~\eqref{14},\qquad {\rm  where}}}
\end{floatequation} 
\begin{eqnarray}
\gamma_1&=&\frac{4}{D-4}[(D-p)(p-\alpha)+\alpha(p+2)-p],\label{15}\\
\gamma_2&=&\frac{1}{D-4}\bigg\{\!(p+1)[(D-p)(p-2\alpha)+6\alpha-p]\!\nonumber\\&&+\!(q\!-\!p)[2\alpha(p-1)\!-\!p(p\!-\!1)\!+\!2p]\bigg\}.\label{16}
\end{eqnarray} In the limit $\alpha_2\rightarrow 0$ Eq. (\ref{14}) recovers the usual case of General Relativity studied in Ref. \cite{COMP}.

Finally, since the internal space is compact, then the integration of the left-hand side of Eq. (\ref{10}) vanishes. Hence the 1-parameter family of consistency conditions is given by 
\begin{eqnarray}
&&\hspace*{-0.7cm}\!\oint W^{\alpha+1}\Bigg\{\alpha\,^{(p+1)}\!RW^{-2}+(p-\alpha)\,^{(D-p-1)}\!R+\gamma_3\,\mr{R}\nonumber\\&&\hspace*{-0.7cm}\!\!+8\pi G_D(1\!-\!2\alpha_2)\!\Bigg[\gamma_1\!-\!a\!+\!\tilde{\gamma}_1\tau_{\mu}^{\;\mu}\!+\!\tilde{\gamma}_2\tau^m_{\;m}\!-\!\gamma_2T_q^{(i)}\tilde\Delta(y\!-\!y_i)\Bigg]\nonumber\\&&+2\alpha_2\!\Bigg[\!\alpha W^{-2}\!A_{GB}\!-\!\gamma_3\mr{R}\!+\!(p\!-\!\alpha)B_{GB}\Bigg]\! \Bigg\}\!=\!0.
\label{17}
\end{eqnarray} In the next section we shall investigate the physical consequences of this general formalism. 

\section{Applications to Thick Branes}

In order to apply the previous formalism, let us consider $D=5$, $p=q=3$. In fact, in a thick brane scenario with one compact extra dimension, the usual effects of the bulk radion  stabilization can be also regarded. Hence we have $\tilde{\gamma}_1=3+2\alpha$, $\tilde{\gamma}_2=4\alpha$, $\gamma_3=3(\alpha+1)$, $\gamma_2=4\tilde{\gamma}_1$, and $\gamma_1=4\gamma_3$. Now, the 1D internal space has a null scalar of curvature $\,^{(1)}\!R=0$ and by taking into account the cosmological data for our Universe, it is fairly reliable to assume $\,^{(4)}\!R$ equals to zero.  Eq. (\ref{17}) thus reduces to 
\begin{eqnarray}
\oint W^{\alpha+1}\Bigg\{8\pi G_5(1\!-\!2\alpha_2)\!\left.\bigg[\!(3+2\alpha)\tau^\mu_{\;\mu}\!-\!\frac{12\Lambda (\alpha+1)}{8\pi G_5}\!\right.\nonumber\\\left.+4\alpha\tau^m_{\;m}\!-\!4(2\alpha+3)\!\sum_i T_3^{(i)}\delta(y-y_i)\bigg]\right.\nonumber\\\left.-2\alpha_2 \bigg[3(\alpha\!+\!1)^{(5)}\!R\!-\!\alpha W^{-2}\!A_{GB}\!+\!(\alpha\!-\!3)B_{GB}\bigg]\!\Bigg\}\!=\!0.\right.\label{18}
\end{eqnarray}
Consequently, the Gauss-Bonnet terms $A_{GB}$ and $B_{GB}$ assume likewise a simpler form. 
For instance, every $D-p-1(=1)$  contribution vanishes. The remaining form is however still far from trivial. Clearly, in the limit $\alpha_2\rightarrow 0$, which does not regard the contribution of $A_{GB}$ and $B_{GB}$, and by considering $\tau^{\mu}_{\;\mu}=0=\tau^m_{\;m}$, it yields 
\begin{eqnarray}
\oint \!W^{\alpha+1}\!\left[12(\alpha+1)\Lambda\!+\!32\pi G_5\!\sum_i T_3^{(i)}\!\delta(y\!-\!y_i)\right]\!=\!0.\label{19}
\end{eqnarray} Thus it leads to $T_3^{(1)}+T_3^{(2)}=0$, for the case $\alpha=-1$ with two branes, evincing a negative brane tension in the Randall-Sundrum model, as expected.  

We note that by disregarding the contribution of $\tau^{\mu}_{\;\mu}$ and $\tau^m_{\;m}$ in the 2-brane Gauss-Bonnet framework the $\alpha=-1$ case is again insightful. The condition to be fulfilled in this case reads 
\begin{eqnarray}  
\!\!\!\frac{\alpha_2}{16\pi G_5}\!\oint\! \Bigg(\!\!4B_{GB}-\!\frac{A_{GB}}{W^2}\!\Bigg)\!=\!(1\!-\!2\alpha_2)\bigg(\!T_3^{(1)}\!\!+\!T_3^{(2)}\!\bigg),\label{20}
\end{eqnarray} showing a consistent model without a negative brane tension. 

Now, thick branes can be obtained from thin branes by replacing the delta source terms by a bulk scalar field. 
Hence, apart from removing the tension terms from Eq. (\ref{18}) it is necessary to particularize $\tau^\mu_{\;\mu}$ and $\tau^m_{\;m}$ as \cite{KGL}
\begin{eqnarray}
\tau^\mu_{\;\mu}\!&=&\!-4\Bigg(\!\frac{1}{2}\Phi'\!\cdot\!\Phi'\!+\!V(\Phi)\!\Bigg),\nonumber\\
\tau^m_{\;m}\!&=&\!\tau^5_{\;5}\,=\,\frac{1}{2}\Phi'\!\cdot\!\Phi'\!-\!V(\Phi),\label{22}\nonumber
\end{eqnarray} where the prime denotes derivative with respect to the extra dimension. On the thick  brane scenario, the $\alpha=-1$ case implies the consistency condition to be:  \vspace{-0.1cm}
\begin{eqnarray}
\alpha_2\!\oint (4B_{GB}\!-\!W^{-2}A_{GB})\!=\!16\pi G_5(1\!-\!2\alpha_2)\!\oint \Phi'\!\cdot\!\Phi'.\label{23}\end{eqnarray}  \vspace{-0.1cm}
Thus, whenever the above left-hand side is positive, it is possible to generate a thick brane  with a periodic extra dimension, circumventing the Gibbons-Kallosh-Linde no-go theorem. Without the Gauss-Bonnet contribution, it does not hold. Obviously, the theorem still holds for the usual case of General Relativity.

In the useful and straightforward 5D case where the metric (\ref{metrica}) is given by 
\begin{eqnarray} W^2(y)\eta_{\mu\nu}dx^\mu dx^\nu+dy^2\,,\end{eqnarray} where $\eta_{\mu\nu}$ denotes the Minkowski 4D metric components, the Einstein-Gauss-Bonnet system can be solved for the warp factor $W(y)$ and a scalar field $\Phi$ with an associated potential. It is important to remark, however, that in a tentative model -- a typical subterfuge used in quite complicated systems, as such one regarded here --- it is possible to envisage a physically interesting warp factor by investigating the output coming from the previous sum rules. In the case specified in this paragraph, Eq. (\ref{23}) implies that 
\begin{eqnarray}
\oint&&\hspace*{-0.6cm}\Bigg\{W^{-10} \left(W^2 \left(4 W^{10}+8 W^6-3 W^4+2
   W^2+1\right) W''^2\right.\nonumber\\&&\left.+2 \left(10 W^8+28 W^6-9 W^4+4
   W^2+2\right) W'^4\right.\nonumber\\&&\left.+2 W \left(4 W^{10}+5 W^8+8
   W^6-3 W^4-4 W^2\right.\right.\nonumber\\&&\left.\left.\qquad\qquad-2\right) W'^2
   W''\right)\Bigg\} > 0\,.
\end{eqnarray} This inequality constitutes thus, an additional physical constraint on the warp factor. 
\section{Final Remarks}
The exact form of $A_{GB}$ and $B_{GB}$ in Eqs.(\ref{11}) and (\ref{12}), respectively, can be calculated  by using the metric (\ref{metrica}). Hence, the integrands in Eqs.(\ref{17}) and (\ref{20}) are shown \emph{not} to be  a total derivative, and thus 
the constraints obtained in Secs. II and III hold, as well as the no-go theorem.
 The impossibility of generating thick branes in a bulk containing a compact internal space performs a strong constraint for braneworld models. It can be circumvented though, in the  context involving the Gauss-Bonnet geometry. We have shown that in this case 
  thick branes are indeed allowed. 
Concerning more physical outcomes from Gauss-Bonnet terms,   similar ones might be achieved in different relativistic theories, as $f(R)$ for instance. These matters are under current investigation \cite{OT,baz}. Finally, in the Gauss-Bonnet context, singular sources are also precluded from the $3$-branes solutions given in Ref. \cite{KM}. 
\begin{center}
*\;*\;*
\end{center}\vspace{-0.3cm}
RdR is grateful to CNPq grants No. 303027/2012-6 and No. 473326/2013-2, FAPESP 2013/10229-5 and CAPES Proc. n$^{o}$ 10942/13-0 for partial financial support.
 JMHS thanks to CNPq (482043/2011-3; 308623/2012-6).

\end{document}